# Explaining Strategic Decisions in Multi-Agent Reinforcement Learning for Aerial Combat Tactics

**Ardian Selmonaj**   **Alessandro Antonucci**
Dalle Molle Institute for Artificial Intelligence (IDSIA) - SUPSI/USI, SWITZERLAND

**Adrian Schneider**   **Michael Rüegsegger**   **Matthias Sommer**
armasuisse Science and Technology, SWITZERLAND

**Abstract.** *Artificial intelligence* (AI) is reshaping strategic planning, with *Multi-Agent Reinforcement Learning* (MARL) enabling coordination among autonomous agents in complex scenarios. However, its practical deployment in sensitive military contexts is constrained by the lack of explainability, which is an essential factor for trust, safety, and alignment with human strategies. This work reviews and assesses current advances in explainability methods for MARL with a focus on simulated air combat scenarios. We proceed by adapting various explainability techniques to different aerial combat scenarios to gain explanatory insights about the model behavior. By linking AI-generated tactics with human-understandable reasoning, we emphasize the need for transparency to ensure reliable deployment and meaningful human-machine interaction. By illuminating the crucial importance of explainability in advancing MARL for operational defense, our work supports not only strategic planning but also the training of military personnel with insightful and comprehensible analyses.

## 1.0   INTRODUCTION

### 1.1   The Potential of AI in Wargames

*Artificial Intelligence* (AI) has become a transformative force across many domains, with major advances in both general applications and specialized areas like wargames. In recent years, AI has shown strong capabilities in strategic decision-making, adaptability, and navigating complex environments, which are key traits for wargaming. A notable branch of AI is *Reinforcement Learning* (RL), where an agent learns an optimal behaviour through trial-and-error interactions with its environment, without needing expert human data to identify effective *Courses of Actions* (CoAs). RL allows autonomous agents to adapt and devise strategies in response to evolving battlefield conditions. Its success in games like chess [1] highlights its potential in high-level reasoning, planning, and execution under uncertainty. *Multi-Agent Reinforcement Learning* (MARL) [2] extends RL to systems with multiple interacting agents in a shared environment. This makes it particularly suitable for wargames, where it can model the complex collaborative and competitive dynamics of military conflicts. MARL supports the simulation of coordinated strategies, real-time adaptation, and emergent behaviours, making it an ideal framework for developing and testing advanced tactics. Integrating MARL into military simulations enables both training and operational planning, offering valuable insights for real-world decision-making. This work focuses on air combat scenarios, where agents are trained using a MARL algorithm. Of particular interest are dogfight scenarios, characterized by close-range, highly manoeuvrable aerial battles between fighter aircraft. Our goal is to train agent pilots and analyse their resulting behaviour.

Various approaches apply RL and MARL to train agents in air combat scenarios. These approaches are not only limited to dogfighting manoeuvres of fighter aircraft, but also include swarms of *Unmanned Aircraft Vehicles* (UAVs) and different types of aircraft (heterogeneous agents). Small-scale aerial combat typically focuses on using RL to *control* the aircraft and gain positional advantage with little risk of return fire. Early methods involved expert systems [4], [5], or hybrids with learning classifiers [6], [7], while more recent work uses RL techniques [8], [9].

To derive stronger CoAs, modern approaches employ more advanced methods such as *Deep Q-Networks* (DQN) [10], *Deep Deterministic Policy Gradients* (DDPG) [11], curriculum learning [12], or self-play, where agents train against copies of themselves [13]. On the other hand, larger engagements focus on high-level tactical decisions [14] or weapon-target assignment [15], i.e., on *planning* of CoAs. Here, MARL is especially suitable due to the curse of dimensionality and the potential to exploit agent symmetries. Advanced methods include multi-agent DDPG [16], hierarchical RL [17], and attention-based neural networks [18]. A previous and advanced method [19] introduced a hierarchical MARL model with an attention mechanism, which can learn to assign importance to specific features, and used *Proximal Policy Optimization* (PPO) [28] for training. The approach also considered heterogeneous agents, which is still rare in current literature. Incorporating heterogeneous agents can increase the complexity of coordination, as agents may be unaware of each other's skills and capabilities. Further approaches for air combat simulations using RL methods can be found in [20].

## 1.2  Understanding Decisions

Deep RL leverages neural networks for decision-making in complex, real-world environments like wargames. However, these models are often perceived as black boxes due to their limited interpretability. *Explainable RL* (XRL) aims to make the decision-making of RL models understandable by revealing why specific actions are taken in given situations. Challenges in XRL include risks related to scientific evaluation and operational reliability, the absence of universally accepted evaluation metrics, and the difficulty of providing comprehensive explanations for complex tasks [3]. Understanding the decision-making process of neural networks is especially crucial in sensitive military operations for diagnosing errors, enhancing model performance, and comprehending intricate agent behaviours. XRL methods play a critical role in building trust among military personnel, ensuring transparency in safety-critical missions, and facilitating compliance with stringent operational and regulatory standards. Furthermore, precise explainability (i.e., meaningful and reliable explanations) contributes to better risk assessment and management, improves coordination between human and AI agents, and supports the integration of advanced AI systems into existing military frameworks while maintaining operational reliability and effectiveness. Air combat scenarios often involve numerous factors, including manoeuvring, targeting, threat avoidance, fuel management, and coordination with other units, where agents must make split-second decisions to achieve strategic objectives. For a concrete example, consider the following scenario: the agent detects an incoming missile from hostile forces. To counter, it quickly releases flares and performs a barrel roll to confuse the missile's heat sensors and evade the enemy's targeting. In this scenario, the observation of the missile acted as a significant feature to perform the action of flare release and barrel roll.

Existing XRL methods [21] can broadly be categorized as: *i*) policy simplification to track decision steps; *ii*) reward influence to decompose the reward signal into sub-components [22]; *iii*) feature contribution to identify most significant information of input data [23]; *iv*) causal models to identify cause-effect relationship; *v*) hierarchical models to inspect sub-policy selection behaviour. As most explainability methods target single-agent RL, we adapt them to the multi-agent setting, following prior work [30]. We review recent advances in MARL explainability and introduce novel use cases highlighting its role in analysing agent behaviour in simulated air combat scenarios (Figure 1). By examining these advancements, we underscore the importance of explainability in understanding and improving agent behaviour, particularly when applied in complex environments like military simulations. Our paper explores how explainability methods can enhance strategic planning, facilitate human-AI collaboration, and ensure the trustworthiness of AI-driven decisions in mission-critical operations. Through these insights, we aim to demonstrate the urgency of explainable MARL for both research and practical deployments in high-stakes scenarios.

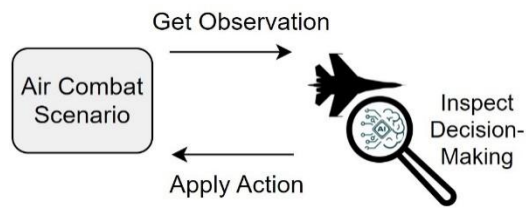

**Figure 1: Overview of inspecting decision-making process.**

## 2.0 LEARNING FRAMEWORKS

The work in this paper primarily involves explainability techniques for air combat simulations trained through a hierarchical MARL framework. We first introduce the concept of MARL and hierarchical RL and afterwards present various XRL techniques and associate them to the domain of simulated aerial battles.

### 2.1 Multi-Agent Reinforcement Learning

MARL is used to solve sequential decision-making problems and involves a set of $N$ agents learning to cooperate or compete through trial and error within a shared environment (Figure 2). The interactions of agents are modelled by a *Partially Observable Markov Game* (POMG) [31], originating from strategic decision-making processes in game theory. With $i \in \{1, 2, \ldots, N\}$ representing the index of each agent, a POMG is defined by the tuple $(S, O_1, \ldots, O_N, A_1, \ldots, A_N, P, R_1, \ldots, R_N, \gamma, \rho_0)$, as follows:

- $S$ is the state-space of the environment.
- $O_i \subset S$ is the (partial) observation space of agent $i$.
- $A_i$ is the action-space for player $i$, yielding the joint action space $\boldsymbol{A} = \times A_i$.
- $P(s'|s, a_1, \ldots, a_N)$ transition probability from state $s$ to state $s'$ when players take actions $a_1, \ldots, a_N$.
- $R_i(s, a_1, \ldots, a_N, s')$ defines the immediate reward for player $i$ when the system transitions from state $s$ to state $s'$ with players taking actions $a_1, \ldots, a_N$.
- $\gamma \in [0,1)$ is the discount factor, used to discount future rewards.
- $\rho_0 \in \Delta(S)$ is the initial state distribution.

Given an initial state $s_0 \sim \rho_0$. At time $t$ in state $s_t \in S$, each agent $i \in N$ gets an observation $o_t \in O_i$ and selects an action $a_i \in A_i$ based on its policy $\pi_i(a \mid o_t)$, which is a probability distribution over actions given observations. After the joint action $\boldsymbol{a}_t = \{a_1, \ldots, a_N\}$ is taken and the game transitions to the new state $s_{t+1}$, each agent receives a reward $R_i(s_t, \boldsymbol{a}_t, s_{t+1})$. The main objective for each agent is to learn a policy $\pi_i^*(a \mid o)$ that maximizes the expected cumulative reward, as follows:

$$\pi_i^* = \underset{\pi_i}{\mathrm{argmax}}\, \mathbb{E}\left[\sum_{t=0}^{\infty} \gamma^t R_i(s_t, \boldsymbol{a}_t, s_{t+1}) \mid \pi_i, \boldsymbol{\pi}_{-i}\right],$$

where $\boldsymbol{\pi}_{-i}$ represents the policies of all agents except $i$. Finding the optimal joint policy $\boldsymbol{\pi}^*$ is challenging because each agent's reward and optimal policy depend on the actions of other agents. This interdependence makes the environment non-stationary for any single agent as other agents' policies evolve. We make use of the *Centralized Training and Decentralized Execution* (CTDE) [27] scheme, the state-of-the-art framework for multi-agent settings. Its popularity is attributed to its capability to address non-stationarity by sharing information amongst agents during training, enhance coordination, and preserve each agent's ability to act independently during execution.

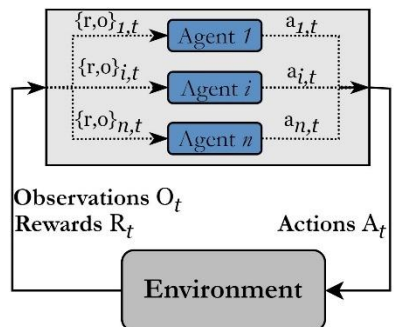

**Figure 2: MARL interaction cycle.**

## 2.2 Hierarchical Reinforcement Learning

To approach the decision structure of defence organization, we extend our MARL model with *Hierarchical Reinforcement Learning* (HRL), which improves learning efficiency by using temporal abstraction to break down tasks into a hierarchy of subtasks. High-level policies $\pi_h$ issue abstract commands to activate low-level policies over a specified time span. The low-level policies $\pi_l$ manage specific actions within a sub-task. This greatly facilitates the training process by exploiting policy symmetries of individual agents and by separating control from command tasks [32]. Combining HRL with MARL yields *Hierarchical Multi-Agent Reinforcement Learning* (HMARL), which is modelled by a *Partially Observable Semi Markov Decision Process* (POSMDP), which was also used in one of our previous works [19]. Our POSMDP includes options, which may last for varying lengths of time and expand the standard concept of actions to include temporally extended CoA. These options can be thought of as macros (commands) taken by $\pi_h$, that consist of multiple primitive (control) actions taken by $\pi_l$. The POSMDP is defined by the tuple $(S, O, A, \mathcal{T}, P, R, T_l, \gamma, \rho_0)$, where:

- $S$ is the state-space of the environment, $O \subset S$ is the observation space and $A$ is the action-space.
- $\mathcal{T}$ is the set of options, where each option $\tau \in \mathcal{T}$ is activated by a high-level action and is defined by a triplet $(I_\tau, \pi_l, \beta_\tau)$:
    - $I_\tau \subseteq S$ initiation set specifies the states from which the option can be initiated.
    - $\pi_l(a|o)$ policy associated with the option, responsible for low-level action selection.
    - $\beta_\tau(s)$ termination condition specifies the probability of the option terminating at state $s$.
- $P(s, \tau, s')$ defines the probability of landing in state $s'$ from state $s$ after the execution of option $\tau$.
- $R(s, \tau, s')$ is the reward function received after transitioning from state $s$ to $s'$ when executing $\tau$.
- $T_l(s, a)$ defines the execution time function, in which an option $\tau$ is active.
- $\gamma \in [0,1)$ is the discount factor, used to discount future rewards.
- $\rho_0 \in \Delta(S)$ is the initial state distribution.

The decision-making process and the objective remains similar to the POMG from Section 2.1. We again use a CTDE approach to train a single high-level commander policy $\pi_h$ for all agents and aircraft types, reflecting policy symmetries where the same strategy applies to any agent. While this makes the commander's behavior homogeneous, individual agents maintain their individual capabilities during missions. Figure 3 shows the relationship between high-level commander policy $\pi_h$ and low-level control policies $\pi_l$. Based on the option $\tau$ chosen by the commander, one of the low-level policies $\pi_l$ is activated per agent for $T_l$ time-steps or until termination condition $\beta_\tau$ is met. When an option terminates, the commander reassesses and determines new tactics. Inspecting low-level policy activations already makes HMARL interpretable, to which we give more details in Section 3.4.

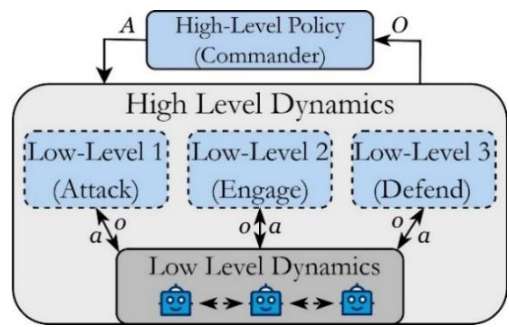

**Figure 3: Hierarchical RL showing high- and low-level dynamics and policies.**

## 3.0 EXPLAINABILITY APPROACHES FOR AIR COMBAT

We now go through the XRL categories [3], [21], and associate them with the multi-agent domain of air combat scenarios to highlight the benefits and indispensability of understanding AI tactics.

The first three methods (policy simplification, reward decomposition, and feature contribution) fall under *reactive* explanations. These focus on short time horizons, offering feedback based on immediate behaviours. For instance, the question "Why did the aircraft fire a missile?" might be answered with "an opponent entered the WEZ". Such explanations emphasise individual actions over broader strategic intent. In contrast, *proactive* explanations span longer time horizons and are better suited for strategic decisions. They might clarify why agents with certain skills switched to defensive mode while others pursued aggressive tactics. Causal and hierarchical RL models support this kind of explanation, offering insights into long-term strategies and coordinated manoeuvres in air combat.

### 3.1 Policy Simplification

In deep RL, neural networks are used as function approximators to learn a decision-making function, namely the policy or the Q-function, where in our analysis, we focus on the policy. Policy simplification refers to the process of reducing the complexity of a policy so that it becomes interpretable by humans. This can be achieved by learning policies in the form of a decision tree to trace each decision step, casting the learned policy as an "if-then" ruleset, using state abstraction to group similar states and reduce the dimensionality of the state space, or representing the learned policy with high-level, human-readable programming languages. The simplicity achieved by these methods is the key advantage, as it facilitates generating explanations and fostering trust in the system.

In simple environments with few agents, policy simplification can generalize well and produce meaningful explanations, which may apply to unexpected air combat scenarios as well. This becomes more difficult in complex settings with multiple objectives and agents of varying skill levels, where explanations often remain static. The primary disadvantage of this approach is the trade-off between model performance and explainability: as the level of explainability increases, the accuracy of the model often decreases. In simulated air combat, where realism is essential for useful insights, maintaining accuracy is critical. This often requires complex models with deep neural networks, careful tuning, advanced algorithms, and highly dynamic scenarios. Although simplification limits policy representation and overall performance, it offers a practical entry point. Simplified policies can train and explain basic control behaviours in air combat to form a solid base for future models that better balance interpretability and performance as scenario complexity grows.

### 3.2 Reward Decomposition

During RL training, the total aggregated reward makes it hard to identify the specific factors influencing decisions. Reward decomposition addresses this by splitting the total reward $R$ into meaningful components $R^i$, each representing a different aspect of the environment. When an agent takes an action $a$ in state $s$, the expected reward is the sum of these components. This decomposition reveals how each $R^i$ contributes to the overall reward, thereby helping to explain the agent's preferences or to identify each agent's contribution towards the team success [24].

To approach this, the Q-function, defined as $Q(s,a) = \mathbb{E}_\pi[R|s,a]$, must also be decomposed with respect to the reward components $R^i$, such that $Q(s,a) = \sum_i Q^i(s,a)$. This is typically achieved by learning separate Q-networks for each $Q^i$. By comparing the differences in Q-values for all actions across each reward component $R^i$, one can gain a deeper understanding of the agent's decision-making process. This allows for two types of analyses:

1) For a given state-action pair $(s,a)$, one can examine the contribution of each component $R^i$ to the overall Q-value. This helps identify which components are driving the agent's action preferences.

2) For a given state and a specific reward component, one can assess which action offers more advantage compared to others. In the case of two actions, computing $\Delta_i(s, a_1, a_2) = Q^i(s, a_1) - Q^i(s, a_2)$ determines the preference of $a_1$ over $a_2$, and vice versa.

Reward decomposition depends on having a reward function that can be meaningfully split into components relevant to the environment and task. Designing such a function is difficult, as it requires incentivizing the right behaviours without causing unintended ones. Poorly designed rewards can lead agents to exploit the function in ways that diverge from the intended goals. In air combat simulations with multiple agents with varying skill levels and changing objectives, defining reward components for each agent or group is essential but complex. Conflicts between components can arise, such as when avoiding radar exposure clashes with the need to engage enemy aircraft. These conflicts can create misalignments between individual and global goals.

Nevertheless, reward decomposition improves transparency by clarifying why an AI agent prefers certain actions or tactics. It helps explain key behaviours like enemy engagement, threat avoidance, and mission success. Commanders can use it to understand, adjust, and validate agent decisions in line with mission objectives. It also allows for adjusting priorities, such as balancing offense and defence. For example, if the agent's reward includes position, distance, fuel use, and hit success, it may choose to retreat because the value of conserving fuel and maintaining position outweighs the reward for scoring a hit. Decomposition reveals that the agent considers fuel efficiency and positioning, not just immediate combat gains. Further, it helps in balancing individual and team goals, such as surviving the mission versus achieving air superiority, therefore supporting more coordinated strategies. Visualization methods like those in [22] further improve the interpretability of agent behaviour in air combat, which we will similarly adapt for our experiments.

### 3.3 Feature Contribution

Closely related to reward decomposition is the method of feature contribution. The objective of this approach is to identify the input features with the greatest influence on decision-making and use them to generate explanations. Techniques in this category include analysing model outputs with modified (contrastive) input data, importance ranking through methods such as saliency maps, assigning Shapley values [24] to each feature or using LIME [25] to approximate the model with a simpler, interpretable model like linear regression. These methods focus on specific input data or local scenarios, but they often fail to explain the complete decision-making process.

Nevertheless, in critical air combat scenarios, these methods can enhance transparency by identifying key environmental features that contribute to tactical success. For instance, in situations where agents are collaborating to attack an enemy aircraft, features like enemy speed or proximity to obstacles can be identified as having the greatest influence on the decision to engage or disengage. By pinpointing such influential features, the training process can be optimized to prioritize the most important tactical considerations, as demonstrated in [23]. In Human-AI collaborative training scenarios, feature contribution methods can help provide and explain recommendations for specific in-game situations, further improving transparency and trust in the AI system.

However, caution must be exercised when interpreting explanations gained from feature contribution or reward decomposition. There is a risk of drawing misguided conclusions. For example, consider a scenario where an agent's decision to retreat is primarily based on the proximity of friendly forces. While this feature may be highlighted as highly influential, it could be only one aspect of a more complex decision-making process. Overemphasizing a single feature can obscure other critical factors, especially in complex air combat simulations with a rapidly changing operational environment. Additionally, scalability and computational overhead are potential limitations of this approach, particularly when applied to large-scale or real-time simulations. To conclude, while feature contribution methods offer valuable insights, they can either enhance or complicate the understanding and performance of MARL systems in dynamic air combat environments. The interpretability provided by these methods must be balanced against their limitations, particularly in highly complex scenarios.

### 3.4 Hierarchical Model

Within HRL, as described in Section 2.2, low-level policies can be analysed independently, which helps in explaining specific decisions within each subtask. Symmetries at lower levels allow the same commands to be reused across similar subtasks, such as operating similar aircraft. Once a sub-policy is understood in one context, its reuse in another makes the behaviour easier to interpret by building on existing knowledge. Hierarchical policies can also be visualized using tools like decision trees or selection-frequency diagrams, providing insight into the sequence of decisions over time. The hierarchical decomposition structure is shown in Figure 4, where high-level commands (options) last for longer periods of time, while low-level actions are executed.

A key limitation is the need to predefine subtasks (options) and low-level policies. This reduces adaptability, especially in unseen environments where new skills must be learned. Still, the hierarchical structure reflects military decision-making structure, which span from high-level planning to low-level manoeuvres. In air combat, where tactical decisions typically account for future outcomes, HRL's temporal abstraction supports long-term planning and aligns with human reasoning.

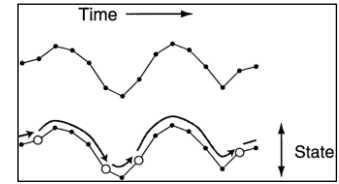

Figure 4: Hierarchical actions last for longer periods of time.

Various subtasks can align with specific mission goals. For example, the high-level objective "achieve air superiority" might be divided into subtasks such as "engage opponent," "evade missile," and "maintain formation." These can be further broken down into basic actions to foster transparency. Reusable sub-policies like "missile evasion" can be applied across different missions. Requiring that one subtask must be completed before another begins, such as allowing "fire missile" only after "engage opponent," reinforces a clear task structure and aids interpretability. However, designing effective hierarchies in air combat demands deep expertise. A deeper hierarchy may enable complex tactics but reduce explainability and a too rigid structure one may fail in performance in unexpected situations. HRL also emphasizes long-term planning, which may lead to actions that appear suboptimal in the short term but serve a broader strategy. Since HRL relies on policies like in standard or multi-agent RL, existing explainability methods can be incorporated to HRL to improve interpretability.

## 3.5 Causal Model

A fundamentally promising approach in XRL is the use of *Structural Causal Models* (SCM). This method captures cause-effect relationships between variables through structural equations. The system distinguishes between endogenous variables $X$ (such as the fuel level of an aircraft) and exogenous variables $Y$ (such as obstacle positions), with equations of the form $X_i = f(PA(X_i), Y_i)$ to capture causal effects, where $PA(X_i)$ are the parent variables of $X_i$. A *Directed Acyclic Graph* (DAG) visually represents these dependencies, with nodes as variables and edges as causal links, as shown in Figure 5. For instance, such visualizations allow to capture situations like the following: agent 1 in state $S_1$ observes an incoming missile with state $S_M$, causing it to evade with action $A_1$ and thereby affecting the state of agent 2 with state $S_2$, further causing agent 2 to reposition with action $A_2$. This example illustrates how SCMs can model the relation of observations to action selection or how state-action pairs influence future states in the transition dynamics $p(s'|s, \boldsymbol{a})$, as is done for example in [29]. However, as the number of observations and agents increases, modelling these relationships becomes increasingly difficult, especially in continuous action and state spaces. Therefore, SCMs are better suited to smaller air combat configurations (e.g., 2-vs-2 scenarios) where meaningful cause-effect relationships can be derived between informational data, agents, and environmental factors. SCMs further allow counterfactual reasoning, similar to feature attribution methods. One could ask, "What would have happened if the situation had been different?" For example, one might explore how enemy tactics would change under a different formation. While building SCMs in multi-agent systems is difficult, a well-designed model can support military decision-making by offering tactical insights.

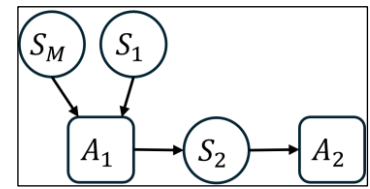

Figure 5: Example of a DAG.

A major limitation of SCMs is their computationally-demanding causal inference in real-time simulations, which is critical in fast-paced aerial combat. Additionally, modelling causal relations in highly complex scenarios is often challenging. The result of a manoeuvre may depend on pilot skill, enemy actions, and environmental factors, which are hard to model precisely. To improve SCM performance in multi-agent air combat, individual SCMs can be created for each agent. These models can reveal how agents influence one another, potentially indicating cooperation tendencies. SCMs can also model hierarchical policies, tracing the effects of tactical commands on lower-level actions and skills. This helps identify when and why certain policies are activated or which conditions activate specific manoeuvres.

## 4.0 EXPERIMENTS AND RESULTS

To explain the tactical decision-making process, we first train air combat agents together with a high-level commander. Once the combat policies have been trained and demonstrate good performance, we use them to generate explanatory insights from various situations using the methods described in the previous Section.

### 4.1 Air Combat Training

Within the domain of aerial battles, we specifically focus on simulated dogfight scenarios[1]. The agents are trained on a dedicated 2D simulation platform developed in Python, with a visualization of the environment shown in Figure 6. We use PPO algorithm [28] for training both high- and low-level agents.

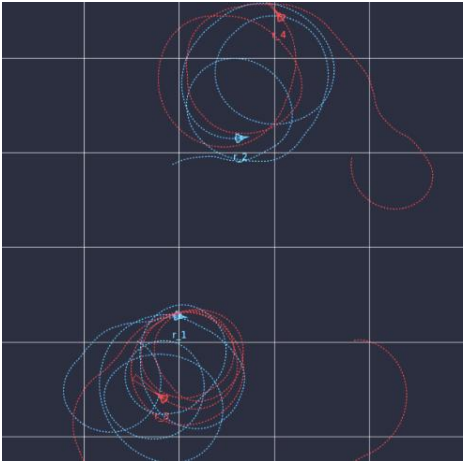

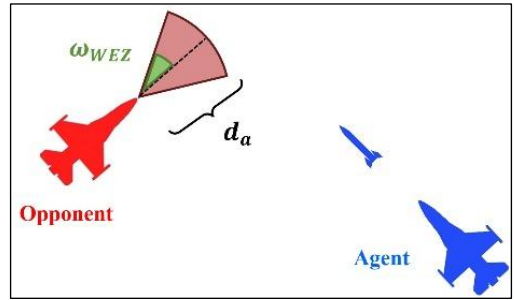

**Figure 6: Visualization of simulation environment.**   **Figure 7: Aircraft attacking mechanism. Blue aircraft are RL agents, red aircraft denote opponents.**

We base our modelling of aircraft on the dynamics of Dassault Rafale[2]. We operate under the assumption of constant altitude within a two-dimensional plane. By restricting the model to 2D, we eliminate the complexities associated with vertical motion, while still preserving key aspects of air combat scenarios. Further, there are *beyond* and *within* visual range air combat scenarios, where we focus here on the latter. To reflect the diversity found in real-world air combat, our approach includes two aircraft types with distinct performance profiles, rendering our model heterogeneous. The first aircraft (AC1) is highly manoeuvrable and armed with rockets, while the second (AC2) lacks rockets but compensates with a longer-range cannon. The detailed dynamics of both aircraft are summarized in Table 1, and their respective attack capabilities are depicted in Figure 7. The agent can attack using either its cannon, which hits the opponent when they enter the *Weapon Engagement Zone* (WEZ) based on a predefined probability, or by firing a missile (rocket).

| Parameter | Symbol | Unit | AC1 | AC2 |
|---|---|---|---|---|
| Angular Velocity | $\omega_{AC}$ | [°/s] | [0, 5] | [0, 3.6] |
| Velocity | $v_{AC}$ | [kn] | [100, 900] | [100, 600] |
| WEZ | $\omega_{WEZ,AC}$ | [°] | [0, 10] | [0, 7] |
| Range | $d_{a,AC}$ | [km] | [0, 2] | [0, 4.5] |
| Hit Probability | $p_{AC}$ | [%] | 0.70 | 0.85 |

**Table 1.** Two different aircraft dynamics (AC1 and AC2) used in our model.

---

[1] Code publicly available at: https://github.com/IDSIA/hhmarl_2D.

[2] https://dassault-aviation.com/en/defense/rafale.

### 4.1.1. Low-Level Training

In our hierarchical model there are three control modes that are learned by the respective low-level policies, i.e., $\pi_l \in \{\pi_a, \pi_e, \pi_d\}$. The behaviour a policy learns is determined by the reward function $R$ (Section 2.1).

1) Attack ($\pi_a$): aggressive behaviour to fight and destroy opponent. With $c_{max}$ being the maximum cannon capacity, $c_{rem}$ the remaining capacity and $R_d = -1$ for getting destroyed, the reward is:

$$R = \frac{c_{max} - c_{rem}}{c_{max}} + R_d, \qquad R \in [-1,1].$$

2) Engage ($\pi_e$): reach and maintain position to face tail of opponent. With $\alpha_{ATA,o}$ being the normalized Antenna-Train-Angle from opponent to agent, and $\alpha_{AA,o}$ the normalized Aspect Angle, the reward is:

$$R = \alpha_{ATA,o} + \alpha_{AA,o}, \qquad R \in [0,2].$$

3) Defend ($\pi_d$): evade from getting destroyed by opponents while avoiding friendly fire ($R_{fr} = -1$):

$$R = R_d + R_{fr}, \qquad R \in [-2,0].$$

We use different approaches to accelerate low-level policy learning and ensure competitive behaviour. First, we employ shared policies for each manoeuvre mode, which are used by all agents. Specifically, every agent $i \in N$ utilizes $\pi_a$ in attack mode, $\pi_e$ in engage mode, and $\pi_d$ in defense mode, irrespective of the aircraft type (AC1 or AC2). Second, we implement a *league-based self-play* mechanism, which incorporates *curriculum learning* with increasing levels of complexity. Initially, our agents are trained against random opponent manoeuvres. Then, we copy the learned policies to the opponents and continue training the agents until the learning rate converges. Once this is achieved, we proceed to train the high-level commander policy, while the low-level policies remain fixed (i.e., no further training). Third, besides using CTDE, some parts of the neural network parameters are shared across the three low-level policies. Each agent receives an observation $o$ containing its own position, orientation, remaining ammunition, as well as that of its closest friendly aircraft and nearest opponent aircraft. The action parameters (manoeuvres) for the low-level policies are shown in Table 3. We consider 5-vs-5 air combat scenarios for training and testing our model. However, by employing shared policies, we emphasize the flexibility of our approach, allowing any combat configuration during inference, such as a 4-vs-2 configuration.

| Parameter | Value |
| --- | --- |
| Relative heading, with $\alpha_h$ the current heading angle | $h \in \{-6, \dots, 6\}, \alpha_h = 15 \cdot h + \alpha_h$ |
| Velocity mapping of $v$ to range of AC1 or AC2 | $v \in \{0, \dots, 8\}$ |
| Shooting with cannon | $c_c \in \{0,1\}$ |
| Shooting with rocket | $c_r \in \{0,1\}$ |

**Table 3**. Control actions of $\pi_a, \pi_e, \pi_d$.

### 4.1.2. High-Level Training

We trained five distinct commander policies $\pi_h$, each with a different sensing capability $m \in [1,5]$. The commander is called for each agent separately and observes the $m$ closest enemy and $m$ closest friendly aircraft and decides which low-level control policy to apply for a specified duration. Specifically, the action space of the commander $\pi_h$ is $A \in \{0,1,2\}$, where 0 activates $\pi_a$, 1 activates $\pi_e$ and 2 activates $\pi_d$ for each aircraft. The total reward consists of killing opponents ($R_k = 1$) and getting destroyed ($R_d = -1$), yielding $R = R_k + R_d$. We also distinguish between *homogeneous* and *heterogeneous* settings. In the homogeneous case, the commander is trained using the more agile AC1 aircraft only. In the heterogeneous case, the commander is trained on randomly formed groups that always include at least one of each aircraft type (AC1 and AC2). In the homogeneous case, all low-level agents behave the same. However, in the heterogeneous case, it is important to know each aircraft's capabilities. Therefore, we incorporate an index to the commander observation to indicate the aircraft type. The combat results (with a commander sensing range of $m = 3$) are shown in Figure 8, while opponents are operating with the same attack and engage policies $\pi_a$ and $\pi_e$ but *without* a commander. As both sides are equipped with the *same* intelligence of low-level policies, we would expect a balanced combat outcome. The results therefore demonstrate that incorporating high-level decision-making significantly improves air combat performance. A win/loss is recorded when all opponents/agents are destroyed, while a draw occurs if at least one aircraft per team remains alive at the end of an episode.

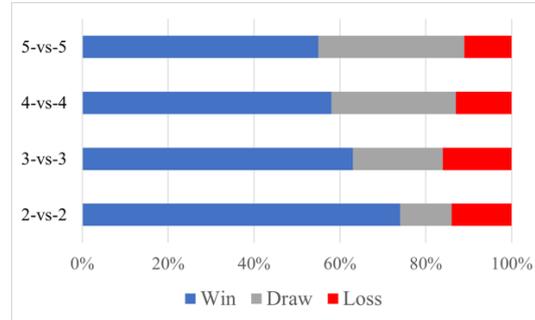

**Figure 8: Combat performance including hierarchal policy $\pi_h$.**

## 4.2 Generating Explanations

Generating explanations for the air combat behaviour of our model is achieved by combining different approaches discussed in Section 3. Given that we employ a hierarchical model, we can utilize the techniques described in Section 3.4, that analyses the selection behaviour of low-level policies by the upper-level instance. Additionally, we apply the method of feature contributions (Section 3.3) by varying initial conditions (different inputs) to observe the resulting outcomes. Finally, we adopt the reward decomposition approach, as outlined in Section 3.2 and similarly applied in [22]. Specifically, we cast reward types (components) to low-level control modes and analyse which low-level control policy has the most significant impact on specific battlefield conditions, as directed by the high-level commander policy. This analysis can be done in two ways: *i) Global*, by examining the overall decisions of all agents involved, and *ii) Local*, by focusing on the decisions of a single aircraft in particular situations. Since our primary interest lies in air combat tactics rather than manoeuvre control, we focus on the hierarchical commander policy's behaviour. We gain explanatory insights by analysing the different commander activations (options) directed to each agent based on modified input data and combat configurations. It is worth noting that the user is free to select which features to alter for outcome inspection. For visualization purposes, we limit the selection to three features (dimensions).

## 4.3 Global Explanations

Since we do not wish to simplify the original policy, and capturing the causal effects of multiple interacting agents is an inherently complex and challenging task, we instead adopt a summarised approach that highlights the most significant decisions made across all agents involved. We do so by focusing on the following features:

1) Opponent strategy: setting the enemy aircraft purely to the three distinct modes (attack, engage and defend) or to a mixed variant, where one of the three modes is randomly selected per aircraft.

2) Combat Difference: Based on the 5-vs-5 combat scene, we increase or decrease the number of agents, e.g., -2 means a 3-vs-5 configuration, with 3 agents and 5 opponents. Similarly, 2 means a 7-vs-5 configuration. The ranges are 1-vs-5 up to 10-vs-5.

3) Sensing capability: alter the radar range $m$ of commander to change the number of detected opponents near an agent.

By perturbing the initial values of these features we inspect the outcomes of the commander to gain an understanding what the crucial inputs are for the decision-making process. As the commander is responsible to decide which low-level policy to activate per agent, we plot the control mode with the highest activation frequency across all feature configurations in Figure 9 for the homogeneous case and in Figure 10 for the heterogeneous one, where each feature corresponds to an axis. To have consistent results, we run 100 simulation episodes per feature combination, where one episode has up to 500 low-level timesteps and up to 25 high-level timesteps (commander decisions). There can be less timesteps, as the episode ends when all agents of one team are destroyed. Additionally, other initial conditions such as positions and angles are randomly sampled and might therefore influence the results.

We first consider the *homogeneous* case with only AC1 aircraft in Figure 9. With a short-sighted sensing capability ($m \leq 3$), the commander tends to neglect long term consequences as it always perceives a restricted number of the total number of opponents. It can therefore less reliably make decisions with respect to the overall game scenario, as it primarily decides for attacking CoAs even in disadvantageous scenarios with less agents than opponents (negative combat difference). This, however, immediately changes when increasing the radar range to $m > 3$, which reveals a more cautious behaviour, where defending and engaging tactics are preferred. Further, with increased radar range and in purely disadvantageous conditions, the commander reliably decides for evading tactics, especially with attacking and mixed opponent configurations. We further notice a rather aggressive behaviour when enemies are set to engaging and defending mode, where the opponents' task is not to fire but rather to flee or gain strategic positioning. In that case, the commander is taking more risk in aiming to destroy opponents. Overall, we may conclude that the sensing capability $m$ of the commander has a high impact on the decision-making process. However, an increasing radar range also led to an overall weaker commander performance. This is because there might be situations where some opponents are far away s.th. they do not influence the current decision significantly and may act as noise to the neural network during training. From the current analysis however, it might still be beneficial to accept a performance decrease to gain a better overview of the combat situation and understand of the model behaviour. Striking the balance between model performance and explainability is a general issue in XRL [3].

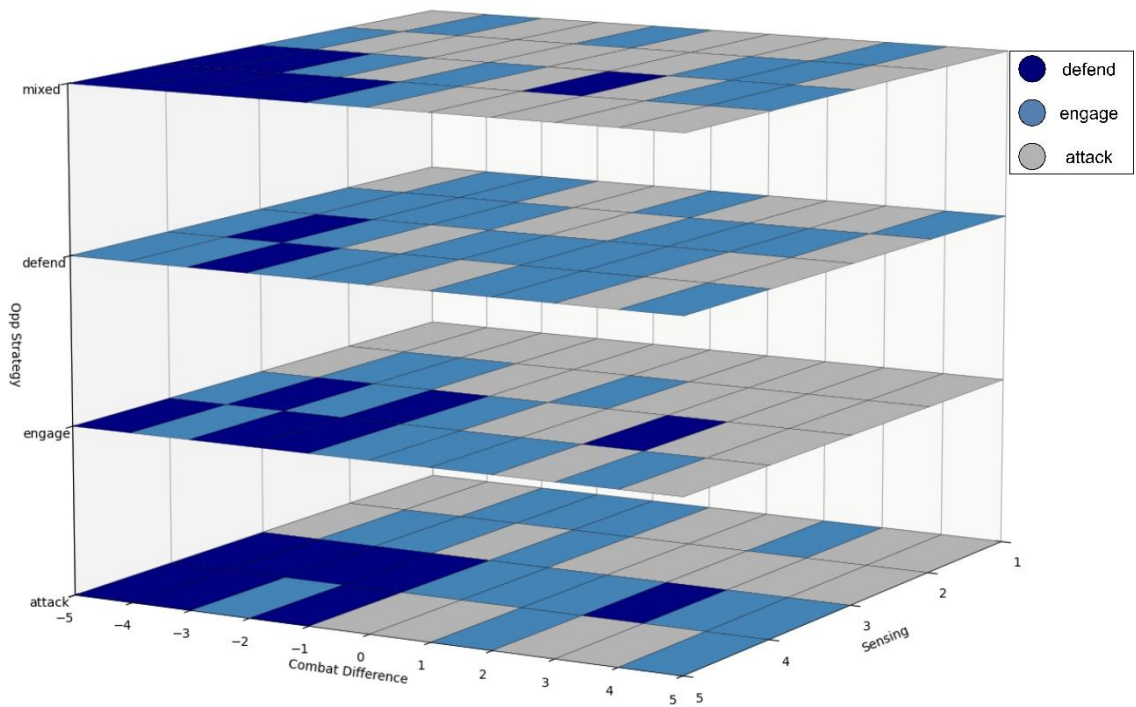

**Figure 9: Homogeneous hierarchy activations *per group*. The mode with the highest frequency is shown.**

We now turn to the *heterogeneous* scenario shown in Figure 10, where both AC1 and AC2 are used for agents and opponents. We compare the resulting behaviours to those observed in the previously analysed homogeneous case. Overall, a similar pattern of tactical decisions remains, with the radar range $m$ still playing a key role in action selection. However, a notable shift can be observed in the frequency and nature of engaging and defensive manoeuvres issued by the commander. The commander takes a more restrained and cautious approach when giving high-risk attack commands, especially when facing opponents with purely aggressive or mixed strategies. This change in behaviour seems to be mainly influenced by the characteristics of AC2, as it has lower manoeuvrability and lacks onboard missiles and therefore faces a clear disadvantage in direct combat. As a result, the commander adjusts its strategy to take into account the higher vulnerability and limited offensive strength of the available air units. From a military perspective, this approach may fit with realistic doctrines in air combat. Commanders are expected to base their decisions not only on the threat situation but also on the specific strengths and weaknesses of their aircraft. In missions where certain units are underpowered or lack offensive equipment, a conservative posture is often adopted to minimise unnecessary losses and to preserve force readiness. These dynamics illustrate the importance of incorporating aircraft heterogeneity into simulations to reflect the complexities and trade-offs inherent in real-world tactical planning.

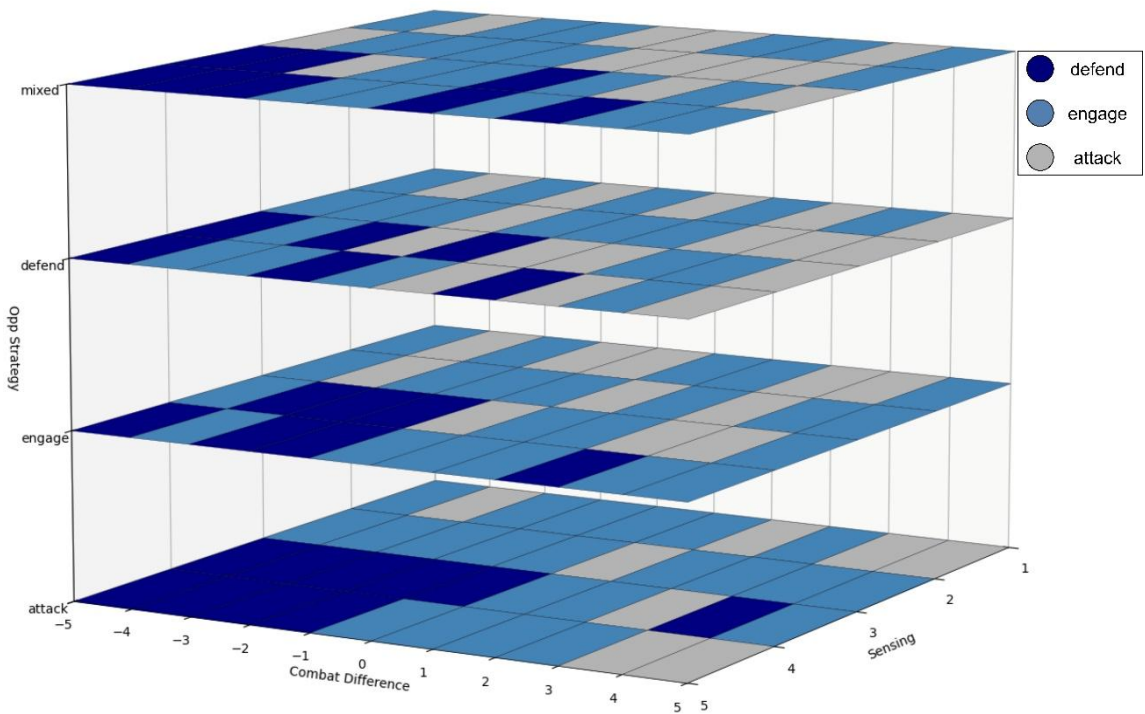

**Figure 10: Heterogeneous hierarchy activations *per group*. The mode with the highest frequency is shown.**

### 4.4 Local Explanations

We now focus on local explanations by considering the hierarchical decisions on one specific aircraft of type AC1. We selected the commander with the best radar range performance ($m = 3$). Once again, we choose three input features to modify:

1) Distance $d$ to closest opponent in kilometres.

2) Antenna Train Angle $\alpha_{ATA}$ to opponent (0°=agent faces opponent, 180°=facing away from opponent).

3) Aspect Angle $\alpha_{AA}$ to opponent (0°=opponent faces away from agent, 180°=opponent faces agent).

As the commander observes the three closest hostile planes and the three closest friendly aircraft per agent ($m = 3$), we only alter the input features of the *nearest* enemy aircraft and set the remaining inputs to random observations (that are still valid). We run 100 experiments per feature combination and plot the low-level mode with the highest activation frequency in Figure 11.

When analysing the local commander decisions for each agent, we observe a consistent pattern across the distance dimension. Specifically, when the agent is in an advantageous position ($\alpha_{ATA} < 60°$ and $\alpha_{AA} < 60°$), the commander consistently favors attacking manoeuvres. However, this behaviour shifts notably when the opponent is facing the agent ($\alpha_{AA} > 120°$), prompting the commander to opt for more evasive actions. From this analysis, we can infer that the Aspect Angle plays a significant role in the commander's decision-making process, as the plots reveal the highest variation along this dimension. The commander expectedly engages opponents when they are at a greater distance, as there is less risk of return fire. This suggests that the decision-making algorithm is effectively balancing offensive and defensive manoeuvres based on both position and proximity. While the results are influenced by the random values generated for other aircraft in the observation, the analysis of 100 simulations per feature combination offers a reliable understanding of the commander's tactical preferences. The consistent behaviour across different configurations reinforces the importance of both Aspect Angle and distance in shaping the commander's decisions, as these features appear to have the highest impact on strategic choices during combat scenarios. This insight highlights the ability of the hierarchical model to adapt its decisions based on evolving battlefield conditions, leading to more effective tactical responses.

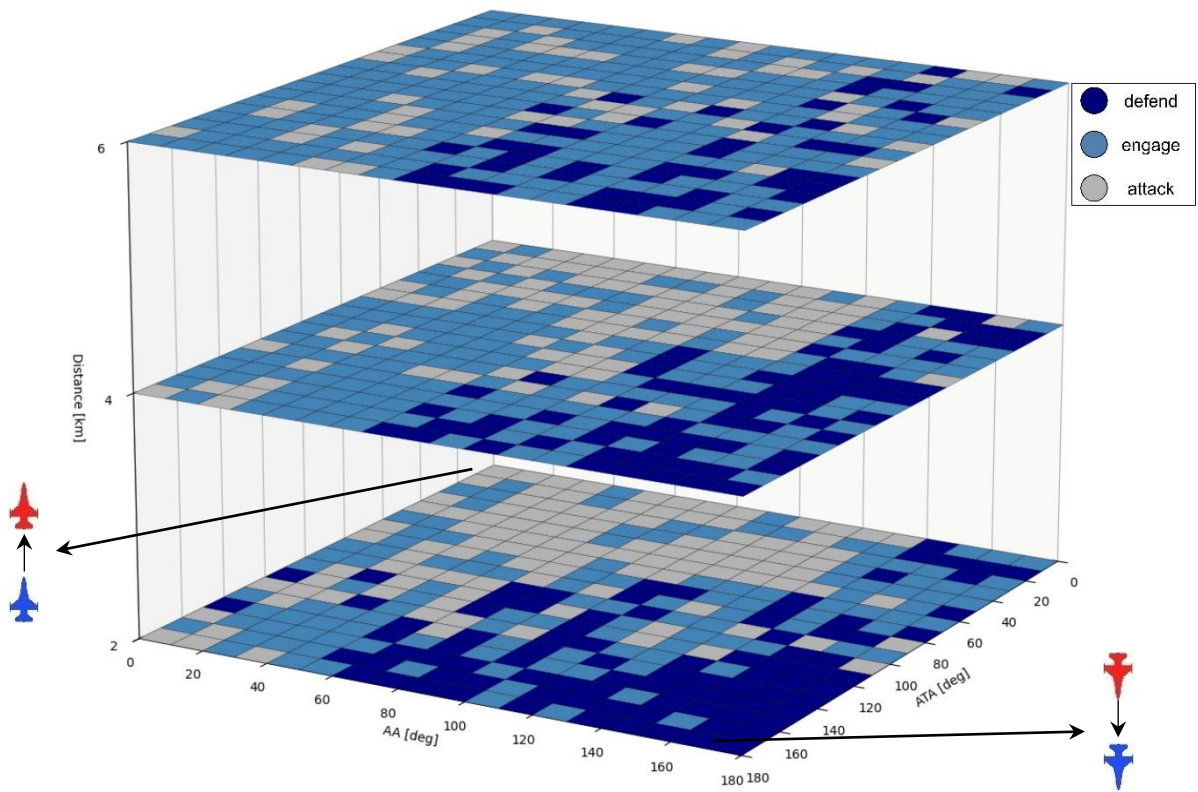

**Figure 11: Hierarchy activations for *one AC1 agent*. The mode with the highest frequency is shown.**

## 5.0 CONCLUSION

In conclusion, while HMARL systems show great potential in improving decision-making processes in air combat scenarios, there is still considerable room for improvement when it comes to explainability. It is important to note that the deployment of AI (and particularly HMARL) in the military domain is still in its early phases, which is even more pronounced regarding explainability tools. A key challenge lies in developing more advanced methods to interpret the internal workings of these AI systems. This would help military personnel understand the reasoning behind complex decisions, which is especially important for building trust and ensuring transparency in high-risk, mission-critical, environments such as air combat.

One fundamental difference between AI systems and human pilots is how decisions are made. AI systems optimise their decision functions (policies) using large volumes of simulated data and pre-defined objectives, processing information at speeds and scales far beyond human capacity. In our experiments, commanders were able to learn different tactical strategies when deploying heterogeneous aircraft dynamics. However, AI systems may still fall short in terms of the intuition and adaptability that human pilots bring to complex and rapidly changing scenarios. Human pilots draw on their experience, intuition, and contextual awareness to respond to unexpected developments. This key difference raises important safety concerns. AI systems might make decisions that are technically optimal according to their model, but operationally unsafe due to a lack of contextual understanding.

Linked to this is the trade-off between performance and explainability as the quality of explanations is tied to the system's performance. In our case, the HMARL model does not yet deliver perfect combat performance. As a result, the explanation method might sometimes produce counterintuitive explanations, for instance, selecting defensive manoeuvres even when in a clearly advantageous position. These behaviours highlight the need to further refine the HMARL model while keeping explainability in focus.

Ensuring that the decision-making process of AI systems is interpretable is essential, as it enables human operators to trust the system and intervene when necessary. Additionally, the information AI systems use to make decisions, such as sensor data, radar signals, and environmental variables may not align with what human pilots consider most relevant. For example, AI may prioritise data-driven metrics like optimal attack angles or speed, while human pilots may rely more on instinct and lived experience. Making these differences transparent is vital for integrating AI systems into human teams and ensuring that humans can effectively and safely use AI techniques during real-world air combat missions.

In summary, making hierarchical MARL more explainable for air combat not only has the potential to improve system performance. It also plays a crucial role in enhancing safety, promoting transparency, and fostering the trust needed for successful deployment in high-stakes environments.

**Acknowledgments**
We gratefully acknowledge Armasuisse S+T for their valuable contribution in contextualizing our research by linking Multi-Agent Reinforcement Learning to real-world military applications. We also thank the NATO Symposium STO-MP-MSG-217 for the opportunity to present our initial findings and engage in constructive discussions.

**Conflict of interest**
The authors declare that there are no financial or non-financial conflicts of interest that could have influenced the research presented in this work.